\documentclass[12pt,a4paper]{article}
\usepackage[utf8]{inputenc}
\usepackage[T2A,T1]{fontenc}
\usepackage[english]{babel}
\usepackage{amssymb,amsfonts,amsmath,mathtext,cite,enumerate,float} %
\usepackage[dvips,pdftex]{graphicx}
\usepackage{geometry} %
\usepackage{ifpdf}
\usepackage{hyperref}

\usepackage{authblk}
\usepackage{blindtext}

\title{Tunneling current between graphene layers}

 \author[*]{Nikolai A. Poklonski}
 \author[*]{Andrei I. Siahlo}
 \author[*]{Sergei A. Vyrko}
 \author[+]{Andrei M. Popov}
 \author[+,$\bot$]{Yurii E. Lozovik}
 \affil[*]{Physics Department, Belarusian State University, pr. Nezavisimosti 4, Minsk 220030, Belarus}
 \affil[+]{Institute of Spectroscopy, Fizicheskaya Str. 5, Troitsk, Moscow Region 142190, Russia}
 \affil[$\bot$]{Moscow Institute of Physics and Technology, Institutskii pereulok 9, Dolgoprudny, Moscow Region 141700, Russia}

\begin{document}
         \maketitle
         \begin{center}
                  Physics Department, Belarusian State University, pr. Nezavisimosti 4, Minsk 220030, Belarus
         \end{center}

\begin{abstract}
The physical model that allows to calculate the values of the tunneling current between graphene layers is proposed. The tunneling current according to the proposed model is proportional to the area of tunneling transition. 
The calculated value of tunneling con-ductivity is in qualitative agreement with experimental data.
\end{abstract}

\section{Introduction}

The theoretical (see, e.g., \cite{Tamura10}, \cite{Bistritzer10}) and experimental \cite{Stetter08} investigations of the tunneling current between carbon layers were provided in the last years.

Tunneling current between carbon layers according to some of theoretical models (see, e.g., \cite{Tamura10}) is a nonlinear function of area of the tunneling contact. In other theoretical models \cite{Bistritzer10} the "finite disorder potential" is introduced to sup-press the divergence of tunneling conductance between the graphene layers. However the known experiments show that the tunneling conductance between nanotube wells \cite{Stetter08} is proportional to the area of the tunneling contact.

In the present paper the model of tunneling current between two graphene layers is proposed. We assume that the time of tunneling transition of an electron between the layers is finite. The proposed model yields the tunneling current proportional to the total area of the tunneling contact and values of the tunneling conductance in qualitative agreement with the experimental data \cite{Stetter08}.

\section{Tunneling current and time of tunneling}

In the proposed model the tunneling current of $\pi $-electrons between two gra-phene layers at electrical potential $U$ between the layers is the sum of the time derivative of probability
of electron tunneling transition from one (top) to an-other (bot) layers 
over all quantum numbers 
$\vec{k}_{top(bot)} = (k_x,k_y)$
(see \cite{Tersoff85})

\begin{equation}\label{eq101}
  I=e \sum_{\vec{k}_{top}}\sum_{\vec{k}_{bot}} \biggl[f \biggl(E_{top},E_F+\frac{eU}{2} \biggr)-f \biggl(E_{bot},E_F-\frac{eU}{2} \biggr) \biggr] \biggl(\frac{\partial w(E_{bot}-E_{top},t)}{\partial t}\biggr)_{t=t_C},
\end{equation}
where $ E_{bot} = E_{bot}(\vec{k}_{bot})$ and $E_{top} = E_{top}(\vec{k}_{top}) $ are the energies of an electron in the two layers, $f(E,E_F) = \{1+ \exp[(E-E_F)/k_BT]\}^{-1}$ is the Fermi function, $E_F$ is the Fermi level, $k_BT$ is the thermal energy.
 The ratio of the probability of electron transition between the states with energies $E_{bot}$ and $E_{top}$ during the time $t_C$ in the first order of perturbation theory is (see, e.g., \cite{Landau03}):

\begin{equation}\label{eq102}
w(E_{bot}-E_{top},t_C)=\frac{2|M_{bot,top}^{\vec{k}_{bot},\vec{k}_{top}}|^2(1-\cos[(E_{bot}-E_{top})t_C/\hbar])}{(E_{bot}-E_{top})^2},
\end{equation}
where $M_{bot,top}^{\vec{k}_{bot},\vec{k}_{top}}$  is the tunneling matrix element between the states $\Psi_{bot}$ and $\Psi_{top}$ of the two graphene layers. 
At $(E_{bot}-E_{top})t_C/\hbar])<<1$ the tunneling rate $\partial w/\partial t$, and thus the tunneling current is proportional to the transition time $t_C$:

\begin{equation}\label{eq103}
 \biggl(\frac{\partial w(E_{bot}-E_{top},t)}{\partial t}\biggr)_{t=t_C}=\frac{2t_C}{\hbar^2}|M_{bot,top}^{\vec{k}_{bot},\vec{k}_{top}}|^2
\end{equation}

Note that the finiteness of the transition time between the quantum states was assumed in \cite{Poklonski03} at considering of interaction of an electron with impurities in doped semiconductors. Note also that for $t_C\rightarrow\infty$ the formula (\ref{eq103}) is transformed to an expression with delta-function 
$
(\partial w/\partial t)_{t_C\rightarrow\infty}= 
(2\pi /\hbar)|M_{bot,top}^{\vec{k}_{bot},\vec{k}_{top}}|^2\delta\left(E_{bot}-E_{top}\right)
$
and gives area-dependent conductivity in \cite{Tamura10} or leads to divergence of the tunneling conductivity in 
\cite{Bistritzer10}.

To find the time $t_C$ of tunneling transition of an electron between the layers we take into account that the local average speed of the electron along $z$ direction is expressed via the phase of its wave function \cite{Kneubuhl94} as

\begin{equation}\label{eq104}
v_z=\frac{\partial}{\partial z}\left(\frac{\hbar \phi}{m_0}\right)=\frac{\hbar}{m_0}\frac{\delta \phi}{\delta Z},
\end{equation}
where $\delta Z$ is the interlayer distance, $m_0$ is the electron mass. 
According to the Aharonov-Bohm effect the wave function of electron being under the electro-static potential $U$ 
during the time $t_C$ experiences a phase shift \cite{Feinberg63}

\begin{equation}\label{eq105}
\delta\phi=eUt_C/\hbar. 
\end{equation}
From Eqs. (\ref{eq104}) and  (\ref{eq105}) we obtain

\begin{equation}\label{eq106}
t_c^2=m_0\delta Z^2/eU.
\end{equation}

The tunneling matrix element between the states $\Psi_{bot}$ and $\Psi_{top}$ of two graphene layers according to the Bardeen formalism \cite{Bardeen61} has the form:

\begin{equation}\label{eq107}
 M_\text{bot,top}^{\vec{k}_\text{bot},\vec{k}_\text{top}} = \frac{\hbar^2}{2m_0}\int_S (\Psi_\text{bot}^{*} \nabla \Psi_\text{top} - \Psi_\text{top} \nabla \Psi_\text{bot}^{*})\,d\vec{S},
\end{equation}
where $S$ is an arbitrary surface between the graphene layers.

Near the $K$-points ($\vec{K}_j$) of the Brillouin zone the relation 
$M_\text{bot,top}^{\vec{k}_\text{bot},\vec{k}_\text{top}}=
M_\text{bot,top}^{\vec{k}_\text{bot}=\vec{k}_\text{top}=\vec{k}}
\delta\left( \vec{k}_\text{bot},\vec{k}_\text{top} \right)$
takes place
\cite{Poklonski13}
and the sum over $\vec{k}_\text{bot}$ and $\vec{k}_\text{top}$ in Eq. (\ref{eq101}) is replaced by the sum over one $\vec{k}$  
$(\sum_{\vec{k}_{bot}}\sum_{\vec{k}_{top}}|M_\text{bot,top}^{\vec{k}_\text{bot},\vec{k}_\text{top}}|^2\times...\rightarrow
\sum_{\vec{k}_{bot}=\vec{k}_{top}=\vec{k}}|M_\text{bot,top}^{\vec{k}_\text{bot}=\vec{k}_\text{top}=\vec{k}}|^2\times...
     )
$. 
Taking this into account Eq. (\ref{eq101}) for the current can be written as:

\begin{equation}\label{eq108}
 I=\frac{2e}{\hbar^2}\sum{k}[f(E_{top},E_F+eU/2)-f(E_{top},E_F+eU/2)]|M_{bot,top}^{\vec{k}}|^2t_C
\end{equation}
Near the $K$-points the matrix element (\ref{eq107}) according to \cite{Poklonski13} is $|M_{bot,top}|=0.11$ eV for the Bernal stacking of graphene layers.

Next, we replace the sum in Eq. (\ref{eq108}) by the integral  
$\sum_{\vec{k}}\textrm{(...)}\rightarrow A\int_{E_{min}}^{E_{max}}g_{gr}(E)\textrm{(...)}dE$, 
where $g_{gr}=A^{-1}dN/dE$ is the electron density of states in graphene, $A$ is the area of graphene. The number of the electron states with the wave vectors in the interval $[\vec{k},\vec{k}+d\vec{k}]$ in the graphene layer with the area 
$A$ is $dN=2\cdot 2\cdot(A/(2\pi)^2)dk_xdk_y$.
Here the first factor '2' is due to two spin directions, the second factor '2' is due to two energy minimums in Brillouin zone of graphene (see \cite{CastroNeto09}); the factor $(2\pi)$ is due to the Heisenberg uncertainty principle 
($A\delta p_x\delta p_y=(2\pi\hbar)^2N$).
Near the $K$-points of the Brillouin zone of graphene the electron energy is a linear function of the distance between $\vec{k}$ and the wave vector of $K$-point:
$E(k)=\pm(\gamma a\sqrt{3}/2)\cdot|\vec{k}-\vec{Kj}|$ \cite{CastroNeto09}. It leads to:

\begin{equation}\label{eq109}
g_{gr}(E)=\frac{1}{A}\frac{dN}{dE}=\frac{4}{3\pi\gamma^2a^2}|E|.
\end{equation}
Taking into account Eq. (\ref{eq109}), the expression (\ref{eq108}) for the current between the graphene layers with the contact area $A$ takes the form:

\begin{equation}\label{eq110}
I=\frac{2\pi e}{\hbar}\sqrt{\frac{m}{eU}}\delta Z \cdot A\cdot\left(\frac{8}{3\pi\gamma^2}\right)
\cdot|M_\text{bot,top}|^2\cdot
(eU)^2\cdot F\left(\frac{eU}{k_BT}\right),
\end{equation}
where

\[
F(eU/k_BT)=
\left(
\frac{k_BT}{eU}\right)^2
\int_{0}^{+\infty}
\frac{2(E/k_BT)\sinh(eU/2k_BT)}{\cosh(eU/2k_BT)+\cosh(E/k_BT)}d(E/k_BT)\approx
\]

\[
\approx\left\{ \begin{array}{ll}
2\ln(2)\cdot(k_BT/eU) & \textrm{at   }eU/k_BT<<4\ln(2)\\
1/4 & \textrm{at   }eU/k_BT>>4\ln(2)
\end{array} \right.
\]
It is seen from Eq. (\ref{eq109}) that the current is proportional to $U^{3/2}$ 
at $eU/(2kBT)<<8\ln2$ 
and to $U^{1/2}$ at $eU/(2kBT)>>8\ln2$.
In contrast to the models \cite{Tamura10,Bistritzer10} the current is proportional to the area of tunneling contact and does not require an "disorder potential" to suppress the divergence of the tunneling conductance.

Eq. (\ref{eq109}) gives the value of the current density $I/A=7.33\cdot10^6 \textrm{A/cm}^2$ at $U=1\textrm{ mV}$, $\delta Z=0.34\textrm{ nm}$ and $k_BT=0.026\textrm{ eV}$, that corresponds to the conductivity 
$g_{\perp}=(I/A)\cdot\delta Z/U=2.5\cdot10^4\space \Omega^{-1}\cdot m^{-1}$. 
The calculated tunneling conductivity is consistent with 
$g_{\perp}=(0.1 \textrm{k}\Omega)^{-1}/\mu\textrm{m}^{-1}=10^4\Omega^{-1}\cdot\textrm{m}^{-1}$, which is measured between the outer walls of multi-walled carbon nanotube \cite{Stetter08}.

\section{Conclusion}

The model of tunneling current between the graphene layers that considers the finiteness of the tunneling transition time is proposed. According to the proposed model the tunneling current is proportional to the area of tunneling contact. The calculated value of tunneling conductivity is in qualitative agreement with the experimental data.

\section*{Acknowledgements} 
Work was partially supported by the 
BFBR (grant Nos. F12R-178, F11V-001) 
and by the RFBR (grant 12-02-90041-Bel).

\section*{Corresponding Authors}
*E-mail: Poklonski@bsu.by; SiahloA@bsu.by

\end{document}